\documentclass[prl,twocolumn,showpacs,floatfix,nofootinbib,groupaddress,showkeys]{revtex4}
\usepackage{psfrag}
\usepackage{graphics}
\usepackage{epsfig}
\usepackage{amssymb}
\usepackage{amsmath}
\usepackage{bm}

\begin{document}

\title{Fisher Waves in the Strong Noise Limit}
\author{Oskar Hallatschek}
\email{ohallats@gmail.com}
\affiliation{Max Planck Research Group for Biological Physics and Evolutionary Dynamics, Max Planck Institute for Dynamics \& Self-Organization (MPIDS), G\"ottingen, Germany}
\author{K.~S.~Korolev}
\email{papers.korolev@gmail.com}
\affiliation{Department of Physics and FAS Center for Systems Biology, Harvard University, Cambridge, Massachusetts 02138, USA}
\begin{abstract}
 We investigate the effects of strong number fluctuations on traveling waves in the Fisher-Kolmogorov reaction-diffusion system. Our findings are in stark contrast to the commonly used deterministic and weak-noise approximations. We compute the wave velocity in one and two spatial dimensions, for which we find a linear and a square-root dependence of the speed on the particle density. Instead of smooth sigmoidal wave profiles, we observe fronts composed of a few rugged kinks that diffuse, annihilate, and rarely branch; this dynamics leads to power-law tails in the distribution of the front sizes.
\end{abstract}

\date{\today}
\maketitle


Traveling waves are a common phenomenon in many systems that combine diffusion and reaction of particles. Familiar examples
are the combustion fronts running through a premixed reactive gas, or the
expanding fronts of bacterial colonies. The standard model for these
wave-like phenomena is the stochastic
Fisher-Kolmogorov-Petrovsky-Piscounov~(sFKPP) equation, the
simplest non-linear model that blends diffusion, growth and number fluctuations. It has been used widely in population
genetics~\cite{Fisher:FisherWave},
ecology~\cite{Murray:MathematicalBiology},
epidemiology~\cite{Murray:MathematicalBiology}, chemical
kinetics~\cite{Kolmogorov:FKPPEquation}, and recently even in quantum
chromodynamics~\cite{Marquet:QCD}.

Our current understanding of traveling waves in reaction diffusion
systems is shaped by studies of the sFKPP equation that have focused
on the weak-noise regime. The opposite limit of strong noise is
relatively unexplored, yet arguably of equal importance. Not only does
it occur naturally, when the driving forces of traveling waves are
weak~\cite{Wilkins:NoisyPopulations}; it also unravels
fundamentally different and often counterintuitive aspects of the
sFKPP equation. The focus of this letter is on two intimately related
questions of how fast these traveling waves move and what their shape
is.

Recently, the velocity of a one-dimensional wave has been computed in the strong noise limit with the help of a duality between the sFKPP equation and~$A\rightleftharpoons A+A$ reaction-diffusion system~\cite{Doering:FisherWaveWeakSelection}. Here, we treat the growth~(reaction) term as a small perturbation, and construct a more direct and more powerful method to calculate the speed of the wave than the one used in Ref.~\cite{Doering:FisherWaveWeakSelection}. Our technique has three important advantages. First, the perturbation expansion enables us to compute
the wave velocity in two dimensions. We find a square root dependence
of the terminal velocity on the particle density. Second, our approach
allows studying the shape of the front on time and length scales that
are dominated by the noise. We find that, in one dimension, the noise
alone can stabilize the front and limit its diffusive broadening. The
front size distribution exhibits power-law tails due to spontaneous
creation and subsequent annihilation of new transition regions. Third,
the perturbation expansion can be applied to more general models with
polynomial reaction terms of higher order.

Without any loss of generality, we discuss the sFKPP equation from the
point of view of population genetics, where it is easy to interpret,
simulate, and potentially test~\cite{Wilkins:NoisyPopulations}. In
this context, the sFKPP is used to describe how a mutation that
increases the growth rate of its carrier spreads through a homogeneous
population~\cite{Fisher:FisherWave} (see also
Ref.~\cite{Korolev:2009p9229} for a recent review). In one dimension, the relative
abundance~$p(t,x)\in [0,1]$ of a beneficial mutation at time~$t$ and
position~$x$ is described by
\begin{equation}
\label{EsFKPP}
\frac{\partial p}{\partial t}=D\frac{\partial^2 p}{\partial x^2}+ap(1-p)+\sqrt{bp(1-p)}\eta(t,x).
\end{equation}
The diffusion term is due to the short-range migration of the
individuals carrying the mutations. The diffusivity~$D$ is
proportional to the average dispersal distance in one generation. The
reaction term,~$ap(1-p)$, accounts for different fitnesses of the
mutant and the ``wild'' type, and the reaction rate~$a>0$ is the
difference in their growth rates. The last term on the right hand side
of Eq.~(\ref{EsFKPP}) describes the sampling error during
reproduction, and is commonly referred to as genetic drift or number
fluctuations. The strength of the noise~$b>0$ is inversely
proportional to the population density, and the (It\^o) white
noise~$\eta(t,x)$ satisfies the following
condition:~$\langle\eta(t_{1},x_{1})\eta(t_{2},x_{2})\rangle=\delta(t_{1}-t_{2})\delta(x_{1}-x_{2})$,
where~$\delta(\cdot)$ stands for Dirac's delta function.

One of the most important predictions of the sFKPP equation is the
velocity~$v$ of an isolated wave that moves from the left half-space
occupied by mutants into the right half-space occupied by the less fit
wild type. The boundary between the half-spaces is assumed to be sharp
initially,~$p(0,x)=1-\theta(x)$, where~$\theta(x)$ is the Heaviside
step function. In the deterministic limit~($b=0$), the wave acquires a
stationary shape with exponential tales, and the velocity of the front
approaches~$v_{F}=2\sqrt{Da}$~\cite{Kolmogorov:FKPPEquation,Fisher:FisherWave}. Surprisingly,
even weak noise, $b^{2}\ll aD$, gives rise to large corrections to the
velocity,~$v=v_{F}-O[\ln^{-2}(1/b)]$~\cite{Brunet:VelocityShiftCuttoff},
indicating the importance of the noise. Here, we access the strong
noise regime, $b^{2}\gg aD$, by constructing a perturbation expansion
in the reaction rate~$a$, while treating the noise term exactly.

Let us define the instantaneous velocity of the wave as the average growth rate of the mutants,~${v}=\frac{d}{d{t}}\left\langle\int_{-\infty}^{\infty}p({t},{x})d{x}\right\rangle$, which is consistent with the usual definition in the deterministic limit. We then take the time derivative inside the integral and use Eq.~(\ref{EsFKPP}) to eliminate~$\frac{\partial p}{\partial {t}}$; after integrating by parts and averaging, we obtain an alternative expression for the velocity,

\begin{equation}
\label{EvelocityMomentRelation}
{v}=a\int_{-\infty}^{\infty} \left\langle
p({t},{x})[1-p({t},{x})]\right\rangle d{x},
\end{equation}

\noindent which relates the wave speed and the instantaneous wave
profile encoded in the dynamical field~$p({t},{x})$.

Note that, to the first order in~$a$, the instantaneous velocity is
given by~$a I({t})/2$, where the
moment~$I({t})=\int_{-\infty}^{\infty} \left\langle
  2p({t},{x})[1-p({t},{x})]\right\rangle d{x}|_{a=0}$ is evaluated in
the neutral limit $a=0$, when neither the mutant nor the wild type
have a selective advantage. The neutral model is exactly solvable
because the hierarchy of the moment equations
closes~\cite{Korolev:2009p9229,Malecot:Dynamics}. For the purpose of
this paper, it is sufficient to consider only the two-point
correlation function~$H({t},{x}_{1},{x}_{2})\equiv\langle
p({t},{x}_{1})[1-p({t},{x}_{2})] \rangle+\langle
p({t},{x}_{2})[1-p({t},{x}_{1})]\rangle$, which is known in population
genetics as the average spatial
heterozygosity. $H({t},{x}_{1},{x}_{2})$~is the average probability
that, at a time~${t}$, two individuals sampled at~${x}_{1}$ and
${x}_{2}$ carry different genetic variants. The equation of motion
for~$H$ is obtained by differentiating its definition with respect to
time and eliminating~$\frac{\partial p}{\partial {t}}$ with the help
of~Eq.~(\ref{EsFKPP}). Note that the rules of the It\^{o} calculus
must be used to properly account for the effects of the
noise~\cite{Korolev:2009p9229,Malecot:Dynamics}. The
result is

\begin{equation}
\label{EOMH}
\frac{\partial H}{\partial {t}}=D\left(\frac{\partial^{2}}{\partial {x}^{2}_{1}}+\frac{\partial^{2}}{\partial {x}^{2}_{2}}\right)H-b H\delta({x}_{1}-{x}_{2}).
\end{equation}

Equation~(\ref{EOMH}) can be understood intuitively as follows. The
probability of sampling two different genetic variants can be traced
to the initial condition by following the lineages of the sampled
variants backward in time. Then,~$H({t},{x}_{1},{x}_{2})$ changes due
to the diffusion and due to the coalescence of the lineages
represented by the term proportional to the delta function. The
lineages coalesce if the genetic variants have a common ancestor,
which can happen only when the lineages occupy the same point in space
at the same time.

We compute~$H({t},{x}_{1},{x}_{2})$ because it has information about the shapes of the wave front and is related to~$I({t})$, and thus~${v}$, by~$I( t)=\int_{-\infty}^{\infty} H({t},{x},{x})d{x}$. To this end, we introduce new spatial variables,~$\xi=({x}_{1}+{x}_{2})/2$ and~$\chi={x}_{1}-{x}_{2}$, encoding the average position of two sampling points and the distance between them, respectively.  After a Laplace transform in~${t}$ and Fourier transform in~$\xi$, Eq.~(\ref{EOMH}) can then be solved for $\tilde H(s,k,\chi)\equiv\int_{0}^{\infty}d{t}e^{-s{t}}\int_{-\infty}^{\infty}d\xi e^{-ik\xi}H({t},\xi,\chi)$. The solution reads

\begin{equation}
\label{EHLaplaceFourierSolution}
\tilde H(s,k,\chi)=\frac{1}{s+Dk^{2}}\left[\frac{e^{-\sqrt{\frac{s}{2D}+\frac{k^{2}}{4}}|\chi|}}{\frac{b}{4D}+\sqrt{\frac{s}{2D}+\frac{k^{2}}{4}}}+\frac{2}{k}\sin\left(\frac{k|\chi|}{2}\right)\right].
\end{equation}

From the definition of~$I(t)$, it follows that the Laplace transform of~$I({t})$ equals~$\tilde H(s,0,0)$. Performing the inverse Laplace transform, we get~$\lim_{{t}\rightarrow\infty}I({t})=4Db^{-1}$. Equation~(\ref{EvelocityMomentRelation}) then implies that the terminal velocity equals~$2aD/b$ in agreement with Ref.~\cite{Doering:FisherWaveWeakSelection}.

Since only the regions with~$p(t,x)\ne0,1$ contribute to~$I(t)$, the
finite limit of~$I(t)$ as~$t\rightarrow\infty$ may look
counterintuitive because it suggests a finite length of the transition
regions in the neutral front; in other words, the noise term alone
limits the diffusive widening of the front. Qualitatively, this
phenomenon can be understood by noticing that the probability of
reaching local fixation~($p=0$ or~$p=1$) is very large at the tails of
the front, so the stationary shape of the front might be maintained by
a balance of the diffusive spreading of the genetic variants into each
other's territory and their subsequent loss due to number
fluctuations~(genetic drift).

\begin{figure}
\includegraphics[width=\columnwidth]{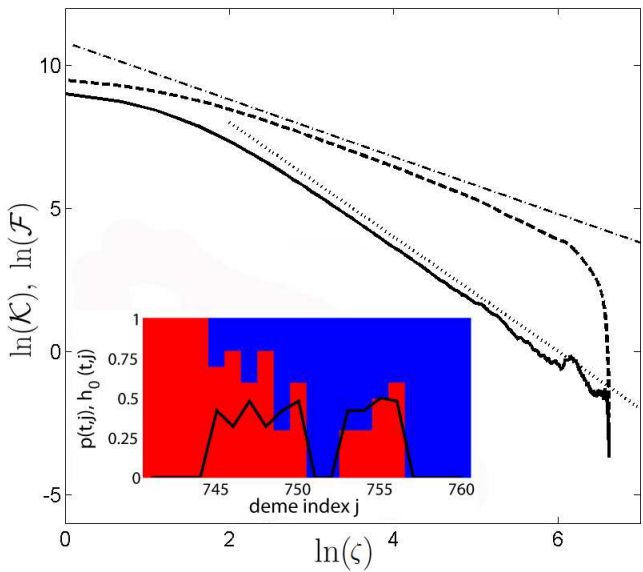}
\caption{(Color online) Simulation of the neutral one-dimensional stepping stone model~\cite{KimuraWeiss:SSM}. The model considers a line of sites~(demes) labeled by an integer~$j$. Each of the demes is occupied by~$N$ individuals that carry one of the two genetic variants. Every generation, neighboring demes exchange $Nm$
migrating organisms. Migration is
followed by Wright-Fisher reproduction, i.e. the next generation is formed by~$N$ organisms sampled from the Bernoulli distribution with the probability of sampling a particular genetic variant equal to its fraction in the current generation. Here,~$N=10$, and~$m=0.1$; the total number of demes is~$1501$. We ensure that the wave front remains within the simulated habitat by moving~$\zeta=0$ to the center of the habitat every third generation. The inset shows a snapshot of the wave front~$p(t,\zeta)$~[red~(darker) color] and the local heterozygosity~$h_{0}(t,\zeta)$~(solid line)
after~$10^{7}$ generations starting from the step function initial condition. The main plot shows~$\mathcal{F}(\zeta)$~(dashed curve) and~$\mathcal{K}(\zeta)$~(solid curve) obtained over~$10^{8}$ generations. For large~$\zeta$, the functions have the following asymptotic scalings~$\mathcal{F}(\zeta)\propto\zeta^{-1}$ and~$\mathcal{K}(\zeta)\propto\zeta^{-2}$ shown by dotted-dashed and dotted lines respectively. }
\label{Fsimulation}
\end{figure}

We analyze this peculiar phenomenon further via particle
simulations. Since our perturbation analysis suggests that, on
sufficiently small length scales, the wave dynamics can be well
approximated by neglecting Darwinian selection, we set $a=0$ in the
simulations. A snapshot of a typical transition region between~$p=1$
and~$p=0$ is shown in the inset of Fig.~\ref{Fsimulation}; the
transition occurs on a very short length scale set by~$D/b$. The inset
also shows the local
heterozygosity~$h_{0}(t,\zeta)=2p(t,\zeta)[1-p(t,\zeta)]$, which is
nonzero only at the kink. The new coordinate~$\zeta$ is defined such
that~$\int_{-\infty}^{0}h_{0}(t,\zeta)d\zeta=\int_{0}^{\infty}h_{0}(t,\zeta)d\zeta$,
i.e. point~$\zeta=0$ is always in the center of the wave. Such a
definition allows us to focus on the shape of the wave by eliminating
the diffusion of the front. It is then instructive to characterize the
average shape of the front
by~$\mathcal{F}(\zeta)=\lim_{t\rightarrow\infty}\langle
p(t,\zeta)\rangle$
and~$\mathcal{K}(\zeta)=\lim_{t\rightarrow\infty}\langle
h_{0}(t,\zeta)\rangle$. In contrast to the narrow boundary shown in
the inset of Fig.~\ref{Fsimulation}, these average characteristics
show power-law tails with exponents close to~$-1$ and~$-2$
respectively; see Fig.~\ref{Fsimulation}.

The power-law tails of~$\mathcal{K}(\zeta)$ and~$\mathcal{F}(\zeta)$
can also be inferred from the exact solution for the two-point
correlation function given by
Eq.~(\ref{EHLaplaceFourierSolution}). One can see that~$\tilde
H(s,k,\chi)=G(s,k)\mathcal{E}(s,k,\chi)$, where~$G=1/(s+Dk^{2})$ is
the diffusion Green's function, which describes the motion of the
center of the wave, and~$\mathcal{E}$ describes the evolution of the
shape of the wave front. This decomposition implies that the front
diffuses with diffusivity~$D$ \textit{independent} of the noise
strength $b$ set by the population density. Furthermore, since
factorization in the Fourier and Laplace domains corresponds to
convolution in the space and time domains, we can think
of~$\mathcal{E}({t}',\xi',\chi)$ as a contribution to~$\tilde
H({t},\xi,\chi)$ from a wave that was present a distance~$\xi-\xi'$
away and time~${t}-{t}'$ ago. 
Power laws in the front size distribution should be reflected in the
asymptotic behavior of $\mathcal{E}({t},\xi,\chi)$. We
get~$\mathcal{E}({t},\xi,\chi=0)=\frac{\Phi({t})}{\sqrt{2\pi{t}}}e^{-\frac{\xi^{2}}{2{t}}}$
from Eq.~(\ref{EHLaplaceFourierSolution}),
where~$\Phi({t})=\int_{-\infty}^{\infty}
d\xi\mathcal{E}({t},\xi,\chi=0)\sim {t}^{-3/2}$ for large times. The
new function~$\Phi({t})$ is the contribution to~$H$ from a transition
region present~${t}$~ago anywhere in space. Since~$\int_{0}^{\infty}
d{t}\mathcal{E}({t},\xi,\chi=0)\sim\xi^{-2}$ for large~$\xi$, we may
conclude that the probability that a kink is located~$\xi$ away from
the center of the wave should decay like a power law with exponent
$-2$. Indeed, this matches the observed behavior of
$\mathcal{K}(\zeta)$ for large arguments.

The algebraic decay of correlations inside a wave front and the slow approach of~$I({t})$ to its limit as~$t\rightarrow\infty$ are in contrast to the narrowness of a typical wave front shown in the inset of Fig.~\ref{Fsimulation}. It is unlikely that a single narrow front relaxes so slowly, but several diffusing fronts can exhibit very slow relaxation, give rise to power-law tails of~$\mathcal{K}$ and~$\mathcal{F}$, and still have a finite value
of~$I({t}=\infty)$. Indeed, we find spontaneous creation of new kinks in our simulations. This process can be interpreted as the following reaction:~$A\rightarrow(2z+1)A$, where~$A$ represents a kink, and~$z$ is an integer. Note that the reaction with~$z=1$ is the most frequent. Neighboring kinks can also merge and subsequently disappear, which is equivalent to the annihilation reaction~$A+A\rightarrow0$. Thus the behavior of the front can be described by the dynamics of Branching Annihilating Random Walks~(BARW)~\cite{Odor:UniversalityClasses}.

Earlier studies of BARW~with an even number of
offspring~\cite{Odor:UniversalityClasses} have found that, in one
dimension, the particles do not proliferate regardless of the birth
rate, which is consistent with the finite value
of~$I({t}=\infty)$. Moreover, we can understand the asymptotic
behavior of~$\mathcal{K}(\zeta)$ and~$\mathcal{F}(\zeta)$ by
considering the dynamics of only three annihilating random
walks~(ARW); higher number of ARW lead to subleading corrections as
one can easily show by generalizing our analysis of three ARW.

The statistical properties of~ARW have been reviewed by Fisher~\cite{Fisher:WWW}. Here, we extend his analysis to compute the average number of ARW present at position~$\zeta$, which is proportional to~$\mathcal{K}(\zeta)$ for large~$\zeta$. Note, from the~ARW interpretation, it follows that~$\mathcal{F}(\zeta)\propto\int\mathcal{K}(\zeta)d\zeta\propto\zeta^{-1}$ because, for large~$\zeta$,~$\mathcal{F}(\zeta)$ is proportional to the probability that the farthest transition region is at least~$\zeta$ away from the origin. So, we only have to calculate~$\mathcal{K}(\zeta)$.

Since three~ARW eventually annihilate, and the process repeats, it is sufficient to consider only one cycle from~$A\rightarrow 3A$ to~$3A\rightarrow A$. Let~$P({t},{x}_{1},{x}_{2},{x}_{3})$ be the probability to find three ARW at time~${t}$ at
positions~${x}_{1}>{x}_{2}>{x}_{3}$, which obeys a three-dimensional diffusion equation, with the following absorbing boundary conditions~$P({t},{x}_{1},{x}_{2},{x}_{2})=P({t},{x}_{1},{x}_{1},{x}_{3})=P({t},{x}_{1},{x}_{2},{x}_{1})=0$. Upon using the solution of this diffusion problem from Ref.~\cite{Fisher:WWW}, we obtain~$\mathcal{K}(\zeta)\propto\int d{t}d{x}_{1}d{x}_{2}d{x}_{3}\delta[\zeta-({x}_{1}-\frac{{x}_{1}+{x}_{2}+{x}_{3}}{3})]P({t},{x}_{1},{x}_{2},{x}_{3})\propto\zeta^{-2}$. Also note that the survival probability of three~ARW~$\int d{x}_{1} d{x}_{2} d{x}_{3} P({t},{x}_{1},{x}_{2},{x}_{3})\sim {t}^{-3/2}$ for long times~\cite{Fisher:WWW}, which matches the behavior of~$\Phi( t)$.
Thus, on large length scales, the one-dimensional sFKPP~equation seems indeed equivalent to a one-dimensional reaction-diffusion system of~BARW.

\begin{figure}
\psfrag{v}{$\frac{v}{v_F}$}
\psfrag{b}{$b=(Nm)^{-1}$}
\psfrag{yin}{$\nu$}
\psfrag{xin}{$Ns$}
\includegraphics[width=\columnwidth]{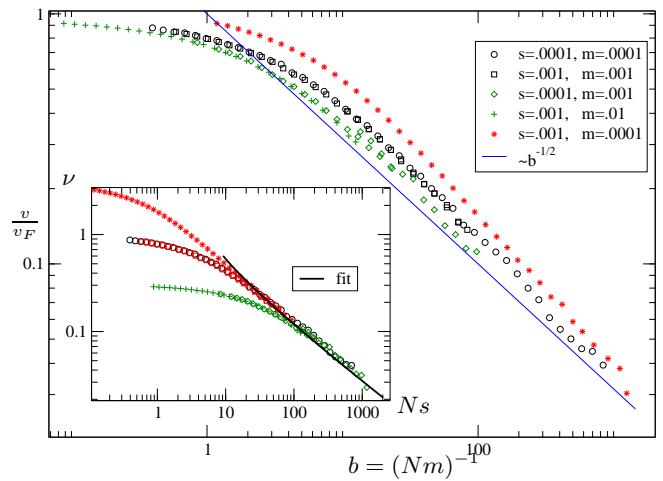}
\caption{(Color online) The speed of planar fronts in the two-dimensional stepping stone model. The simulations are identical to the ones described in Fig.~\ref{Fsimulation} except for two differences. First, we consider a (co-moving) square lattice of $250\times 250$ sites instead of a linear array. Second, a mutant is by a factor $1+a\tau$ more likely to reproduce than a wild type in a single generation time~$\tau$. The simulations are initialized by a sharp step-like front profile and run for $2\times 10^7$ generations.
In the scaling regimes, the wave speeds closely follow a power law $\propto b^{-1/2}=(Nm)^{1/2}$. Deviations are due to log-corrections, and mimic a power law with exponent slightly smaller than $-1/2$. However, the scaling regimes almost collapse in a plot $\nu=(v/v_F)\sqrt{a/m}$ versus $b*m/a=(Ns)^{-1}$, see the inset. The solid line is the solution of
$\nu^2=c_1*Ns/\log[c_2*\nu]$ with the fitting parameters $c_1=0.45$ and $c_2=1$.  }
\label{2dsimulations}
\end{figure}

What happens in higher dimensions? The noisy FKPP equation Eq.
(\ref{EsFKPP}) can easily be extended to two spatial coordinates upon substituting $x\to{\bf r}=(x,y)$, provided one limits the short-wave length variations of the field $p(t,{\bf r})$ by a cutoff $l$, e.g., representing the lattice constant in the model.  The speed of a planar wave front traveling in the $x$ direction again takes the form of Eq.~(\ref{EvelocityMomentRelation}), and depends on the moment $I(t)$. Solving the two-dimensional neutral version of Eq.~(\ref{EsFKPP}) yields 

\begin{equation}
\label{eq:moment-two-dimensions}
\tilde I(s)=\frac{\sqrt{2D} s^{-3/2}}{1+\frac{b}{8\pi D}  \ln\left(\frac{32 \pi^{2} D}{l^2 s}\right) } \qquad \text{(2d)}
\end{equation}

\noindent for the Laplace transform of $I(t)$. Up to logarithmic
corrections, which are typical for two dimensional diffusion problems
\cite{Redner-book2007}, we thus find that the second moment increases
as $I(t)\sim D^{3/2}b^{-1}t^{1/2}$ for long times. Consequently, the
wave speed should increase without bound as $v(t)=aI(t)/2\propto
t^{1/2}$, rendering the perturbation expansion singular. We circumvent
this difficulty by noticing that, when the front is moving, it does
not have enough time to fully relax as predicted by the neutral
dynamics. Relaxation only occurs over a limited time ${t}^{*}$, which
is roughly the time when deterministic motion of the front is of the
same order as its diffusive motion:~${v}{t}^{*}=\sqrt{2D{t}^{*}}$,
i.e.~${t}^{*}=2D{v}^{-2}$. If this time is large, we may impose a
self-consistency condition, $v=2 a I({t}^{*})$, which leads to the
scaling $v\sim D\sqrt{a/b}$ (up to log-corrections) for the wave
velocity. These heuristic arguments, which can be formalized by a
multiple-scale ansatz~\cite{Korolev-halla:Noisy-FKPP-long-version},
thus predict a crossover from the no-noise speed $v\sim 2\sqrt{aD}$ to
$v\sim D\sqrt{a/b}$ for large values of the noise strength, $b\gg
D$. 
Simulated (genetic) wave fronts indeed exhibit this crossover, see Fig.~\ref{2dsimulations}.

In conclusion, we have presented a perturbative approach to the one-dimensional sFKPP reaction-diffusion system with strong number fluctuations, which can be extended to higher dimensions using self-consistency arguments. In one and two dimensions, we found a linear and a novel square-root relationship between the speed of traveling waves and the particle density, respectively. The wave profiles have also been analyzed in one dimension using three different approaches: Simulations, the correspondence between~BARW and sFKPP~equation, and the exact solution for the two-point correlation function. All of these approaches predict that the front size distribution has a power-law tail with a cutoff, which is in contrast to the exponential tails of deterministic Fisher waves. The power-law tail is best understood as a consequence of spontaneous creation of several transition regions that behave as branching annihilating random walks.  Finally, we note that our perturbation expansion also applies to generalized sFKPP equations with higher order polynomial reaction terms. This can be used, for instance, to show that the spread of recessive and dominant beneficial mutations is the same~\cite{Korolev-halla:Noisy-FKPP-long-version}.


\begin{acknowledgments}

\end{acknowledgments}

\bibliography{FisherWavePRL}

\begin{thebibliography}{14}
\expandafter\ifx\csname natexlab\endcsname\relax\def\natexlab#1{#1}\fi
\expandafter\ifx\csname bibnamefont\endcsname\relax
  \def\bibnamefont#1{#1}\fi
\expandafter\ifx\csname bibfnamefont\endcsname\relax
  \def\bibfnamefont#1{#1}\fi
\expandafter\ifx\csname citenamefont\endcsname\relax
  \def\citenamefont#1{#1}\fi
\expandafter\ifx\csname url\endcsname\relax
  \def\url#1{\texttt{#1}}\fi
\expandafter\ifx\csname urlprefix\endcsname\relax\def\urlprefix{URL }\fi
\providecommand{\bibinfo}[2]{#2}
\providecommand{\eprint}[2][]{\url{#2}}

\bibitem[{\citenamefont{Fisher}(1937)}]{Fisher:FisherWave}
\bibinfo{author}{\bibfnamefont{R.}~\bibnamefont{Fisher}},
  \bibinfo{journal}{Ann. Eugenics} \textbf{\bibinfo{volume}{7}},
  \bibinfo{pages}{353} (\bibinfo{year}{1937}).

\bibitem[{\citenamefont{Murray}(2003)}]{Murray:MathematicalBiology}
\bibinfo{author}{\bibfnamefont{J.}~\bibnamefont{Murray}},
  \emph{\bibinfo{title}{{Mathematical Biology}}}
  (\bibinfo{publisher}{Springer}, \bibinfo{year}{2003}).

\bibitem[{\citenamefont{Kolmogorov et~al.}(1937)\citenamefont{Kolmogorov,
  Petrovsky, and Piscounov}}]{Kolmogorov:FKPPEquation}
\bibinfo{author}{\bibfnamefont{A.}~\bibnamefont{Kolmogorov}},
  \bibinfo{author}{\bibfnamefont{N.}~\bibnamefont{Petrovsky}},
  \bibnamefont{and}
  \bibinfo{author}{\bibfnamefont{N.}~\bibnamefont{Piscounov}},
  \bibinfo{journal}{Moscow Univ. Bull. Math.} \textbf{\bibinfo{volume}{1}},
  \bibinfo{pages}{1} (\bibinfo{year}{1937}).

\bibitem[{\citenamefont{Marquet et~al.}(2006)\citenamefont{Marquet, Peschanski,
  and Soyez}}]{Marquet:QCD}
\bibinfo{author}{\bibfnamefont{C.}~\bibnamefont{Marquet}},
  \bibinfo{author}{\bibfnamefont{R.}~\bibnamefont{Peschanski}},
  \bibnamefont{and} \bibinfo{author}{\bibfnamefont{G.}~\bibnamefont{Soyez}},
  \bibinfo{journal}{Phys. Rev. D} \textbf{\bibinfo{volume}{73}},
  \bibinfo{pages}{114005} (\bibinfo{year}{2006}).

\bibitem[{\citenamefont{Wilkins and Wakeley}(2002)}]{Wilkins:NoisyPopulations}
\bibinfo{author}{\bibfnamefont{J.}~\bibnamefont{Wilkins}} \bibnamefont{and}
  \bibinfo{author}{\bibfnamefont{J.}~\bibnamefont{Wakeley}},
  \bibinfo{journal}{Genetics} \textbf{\bibinfo{volume}{161}},
  \bibinfo{pages}{873} (\bibinfo{year}{2002}).

\bibitem[{\citenamefont{Doering et~al.}(2003)\citenamefont{Doering, Mueller,
  and Smereka}}]{Doering:FisherWaveWeakSelection}
\bibinfo{author}{\bibfnamefont{C.}~\bibnamefont{Doering}},
  \bibinfo{author}{\bibfnamefont{C.}~\bibnamefont{Mueller}}, \bibnamefont{and}
  \bibinfo{author}{\bibfnamefont{P.}~\bibnamefont{Smereka}},
  \bibinfo{journal}{Physica} \textbf{\bibinfo{volume}{A325}},
  \bibinfo{pages}{243} (\bibinfo{year}{2003}).

\bibitem[{\citenamefont{{K. S. Korolev} et~al.}(2009)\citenamefont{{K. S.
  Korolev}, Avlund, Hallatschek, and {David R. Nelson}}}]{Korolev:2009p9229}
\bibinfo{author}{\bibnamefont{{K. S. Korolev}}},
  \bibinfo{author}{\bibfnamefont{M.}~\bibnamefont{Avlund}},
  \bibinfo{author}{\bibfnamefont{O.}~\bibnamefont{Hallatschek}},
  \bibnamefont{and} \bibinfo{author}{\bibnamefont{{David R. Nelson}}},
  \bibinfo{journal}{http://arxiv.org/abs/0904.4625v1}  (\bibinfo{year}{2009}).

\bibitem[{\citenamefont{Brunet and
  Derrida}(1997)}]{Brunet:VelocityShiftCuttoff}
\bibinfo{author}{\bibfnamefont{E.}~\bibnamefont{Brunet}} \bibnamefont{and}
  \bibinfo{author}{\bibfnamefont{B.}~\bibnamefont{Derrida}},
  \bibinfo{journal}{Phys. Rev. E} \textbf{\bibinfo{volume}{56}},
  \bibinfo{pages}{2597} (\bibinfo{year}{1997}).

\bibitem[{\citenamefont{Mal{\'e}cot}(1975)}]{Malecot:Dynamics}
\bibinfo{author}{\bibfnamefont{G.}~\bibnamefont{Mal{\'e}cot}},
  \bibinfo{journal}{Theor. Popul. Biol.} \textbf{\bibinfo{volume}{8}},
  \bibinfo{pages}{212} (\bibinfo{year}{1975}).

\bibitem[{\citenamefont{Kimura and Weiss}(1964)}]{KimuraWeiss:SSM}
\bibinfo{author}{\bibfnamefont{M.}~\bibnamefont{Kimura}} \bibnamefont{and}
  \bibinfo{author}{\bibfnamefont{G.}~\bibnamefont{Weiss}},
  \bibinfo{journal}{Genetics} \textbf{\bibinfo{volume}{49}},
  \bibinfo{pages}{561} (\bibinfo{year}{1964}).

\bibitem[{\citenamefont{{\'O}dor}(2004)}]{Odor:UniversalityClasses}
\bibinfo{author}{\bibfnamefont{G.}~\bibnamefont{{\'O}dor}},
  \bibinfo{journal}{Rev. Mod. Phys.} \textbf{\bibinfo{volume}{76}},
  \bibinfo{pages}{663} (\bibinfo{year}{2004}).

\bibitem[{\citenamefont{Fisher}(1984)}]{Fisher:WWW}
\bibinfo{author}{\bibfnamefont{M.}~\bibnamefont{Fisher}}, \bibinfo{journal}{J.
  Stat. Phys.} \textbf{\bibinfo{volume}{34}}, \bibinfo{pages}{667}
  (\bibinfo{year}{1984}).

\bibitem[{\citenamefont{Redner}(2007)}]{Redner-book2007}
\bibinfo{author}{\bibfnamefont{S.}~\bibnamefont{Redner}},
  \emph{\bibinfo{title}{A Guide to First-Passage Processes}}
  (\bibinfo{publisher}{Cambridge University Press}, \bibinfo{year}{2007}).

\bibitem[{\citenamefont{Korolev and
  Hallatschek}()}]{Korolev-halla:Noisy-FKPP-long-version}
\bibinfo{author}{\bibfnamefont{K.~S.} \bibnamefont{Korolev}} \bibnamefont{and}
  \bibinfo{author}{\bibfnamefont{O.}~\bibnamefont{Hallatschek}},
  \bibinfo{note}{unpublished}.

\end{thebibliography}
\bibliographystyle{apsrev}

\end{document}